\documentclass[preprint,5p]{elsarticle}

\usepackage{amssymb}

\usepackage[T1]{fontenc}
\usepackage{dcolumn}
\usepackage{bm}
\usepackage{hyperref}

\usepackage{type1cm}

\journal{Physics Letters A}

\begin{document}

\begin{frontmatter}



\title{On the Empirical Relevance of the Transient in Opinion Models}


\author[1,3]{Sven Banisch}
\ead{sven.banisch@universecity.de}
\author[2,3]{Tanya Ara\'{u}jo}
\ead{tanya@iseg.utl.pt}
\address[1]{Mathematical Physics, Physics Department, Bielefeld University, 33501 Bielefeld (GERMANY).}
\address[2]{Research Unit on Complexity in Economics (UECE), ISEG, TULisbon, 1249-078 Lisbon (PORTUGAL).}
\address[3]{Institute for Complexity Science (ICC), 1249-078 Lisbon (PORTUGAL).}

\begin{abstract}
While the number and variety of models to explain opinion exchange dynamics is huge, attempts to justify the model results using empirical data are relatively rare.
As linking to real data is essential for establishing model credibility, this Letter develops a empirical confirmation experiment by which an opinion model is related to real election data. 
The model is based on a representation of opinions as a vector of $k$ bits.
Individuals interact according to the principle that similarity leads to interaction and interaction leads to still more similarity.
In the comparison to real data we concentrate on the transient opinion profiles that form during the dynamic process.
An artificial election procedure is introduced which allows to relate transient opinion configurations to the electoral performance of candidates for which data is available.
The election procedure based on the well--established principle of proximity voting is repeatedly performed during the transient period and remarkable statistical agreement with the empirical data is observed.
\end{abstract}


\end{frontmatter}


\section{Introduction}

Using physical tools in the analysis of social collective phenomena can help uncovering invisible structures, patterns and mechanisms at work in real--world social systems.
The work of Fortunato and Castellano in 2007 (\cite{Fortunato2007}) is at the leading edge of this endeavour. 
Their analysis of the electoral performance of candidates in various proportional elections revealed an universal voting pattern, which was shown to be independent of the characteristics of the voting population, being instead, a consequence of the elementary interactions.

Empirical data coming from electoral contexts provides one of the most relevant accounts of preference distributions in existing societies. 
An opinion model with empirical relevance should match these accounts of real--world preference distributions.

In a recent paper (\cite{Banisch2009}), we analysed the interplay of opinion dynamics and communication networks. 
Using a bit--string model it was shown that non--trivial social structures emerge from simple rules for individual communication.
Here, using the same abstract bit--string model, we show that the universal scaling function found in Ref.\cite{Fortunato2007} is reproduced when artificial elections are run on the transient opinion profiles. 
Such an empirical confirmation further increases our confidence on the model capabilities to capture and to reproduce some fundamental aspects of real--world dynamics of opinion exchanges.

\section{Opinion Models and Election Data}

Attempts to compare model results to empirical data are relatively rare in opinion dynamics (\cite{Sobkowicz2009}).
There are however some studies with a reference to real data which mostly use election results in the comparison (see~\cite{Castellano2009}, Sec.III.H for an overview).
This was initiated by a statistical analysis of the 1998 Brazilian elections by Filho et al. (\cite{CostaFilho1999}) which revealed that the distribution of votes among candidates ($P(v)$) follows a hyperbolic law (i.e., $P(v) \propto \frac{1}{v}$) in a range of two orders of magnitude.
Similar patterns were found for the Indian elections (\cite{Gonzalez2004}).
However, due to party commitment or strategic voting behaviour a universal scaling could not be expected (\cite{Fortunato2007,Castellano2009}).

A different scenario characterizes the so--called Proportional Elections, where each party competes with an open list of candidates for multiple seats in the parliament.
In Ref.~\cite{Fortunato2007}, the statistical analysis of proportional elections in Italy (1958, 1972, 1987), Poland (2005) and Finland (2003) revealed that "the distribution of the number of votes received by the candidates is a universal scaling function, identical in different countries and years" (p. 1).
This remarkable result is obtained by a re--scaling of the vote numbers $v$ by the number of candidates of the same party $Q$ and the total number of votes received by this party $N$.
The distribution of the function $F(\frac{vQ}{N})$ is the same for all the elections considered and a log--normal fit is shown to approximate the data quite well.

Opinion studies referring to these new empirical insights either concentrate on adaptations of the Sznjad model (\cite{Bernardes2002,Gonzalez2004,Sousa2005}) or on very simple models of opinion spread in different network topologies (\cite{Travieso2006,Fortunato2007}).
Using a Sznjad model variant (see Refs. \cite{Sznajd-Weron2000,Sznajd-Weron2004}) where the opinion states directly account for the preference for one out of a set of candidates, Bernardes et al. (\cite{Bernardes2002}) show that a microscopic opinion model reproduces the characteristic $\frac{1}{v}$--pattern of the 1998 Brazilian elections.
In this approach, there is first a stage to construct an adequate initial condition, in which different candidates have different initial chances of being voted, and secondly a stage in which the usual Sznajd process is performed in order to represent the electoral campaign.
The latter dynamical process is stopped at some (arbitrary) iteration number and the respective transient state is used in the comparison to real--world results.

Subsequent studies (\cite{Gonzalez2004,Sousa2005}) basically use the same mechanisms and analyse the effects of different network structures on the distribution of votes.
The actual problem with the approach due to Bernardes and colleagues (\cite{Bernardes2002}) is the termination of the Sznajd process after a "certain carefully chosen time" (\cite{Castellano2009}, p. 612). 
No reasonable argument is presented for the choice of this iteration number. 
Furthermore, it is not entirely clear how much of the similarity is due to the quite complicated construction of the initial condition.

An alternative opinion model capable of reproducing the voting pattern of the Brazilian and the Indian elections was proposed by Travieso and Costa in 2006 (\cite{Travieso2006}).
Voters are treated as the nodes of a network. 
Initially, some of these nodes are assigned to a favourite candidate and all the others are treated as undecided.
Then, decided nodes are chosen randomly and all their undecided neighbours are associated to the respective candidate.
Already decided nodes change the candidate preference with a given switching probability.
In some sense, this model is similar to the set--up stage of the initial conditions in Ref.~\cite{Bernardes2002}.
This simple model is run on Erd\~{o}s--R\'{e}nyi and Barab\'{a}si networks and it succeeds in reproducing the pattern in the first but not in the latter case.
In Ref.~\cite{Fortunato2007}, a similar model of opinion spread on treelike graphs is used to explain the universal pattern found for proportional elections.

In what follows, an alternative microscopic explanation is provided.  
Briefly,  the essential lines used in previous attempts to empirically confirm opinion models are the following:
 \emph{(i)} the models attempt to reproduce universal patterns, which are normally expressed in terms of scaling laws; 
 \emph{(ii)} in explaining these universal patterns, the system underlying topology is frequently called into place and
  \emph{(iii)} the dynamical process comprise three different periods in time: $(1) $ setting up initial conditions,  
  $(2) $ the process final (steady) state and  $(3) $ an intermediate time interval lying in between the  $(1) $ and  $(2)$.

\section{Method}

\subsection{The Model}

In our recently introduced model of opinion exchange (Ref. \cite{Banisch2009}), opinions are represented as a series of $k$ bits, accounting for the positions concerning $k$ different issues in the agents mind. 
This is similar to the well--known model of cultural dissemination introduced by Axelrod (\cite{Axelrod1997,Castellano2000}).
In the beginning of the simulation $N$ agents are generated and a random bit--string is assigned to them. 
In the iteration process, two agents meet at random.
The two players ($i,j$) are willing to communicate about an issue (one element of the bit--string), only if the number of unequal bits is below or equal to a similarity threshold $d_I$ (i.e., $d(x_i, x_{j}) \leq d_I$).
The result of successful communication is that the agent chosen first ($x_i$) adopts the opinion of the other ($x_{j}$) by flipping one of the unequal bits.
The conceptual idea behind this is that provided that the views of two individuals are close enough, similarity leads to interaction and interaction leads to still more similarity. 
These dynamic rules are summarized in the following steps:
\begin{enumerate}
	\item
	An initial random set--up of $N$ bit--strings of length $k$ according to the uniform distribution;
	\item
	a dynamic process which iterates:
		\begin{enumerate}
		\item
		random choice of two agent strings $x_i, x_{j}$,
		\item
		compute the Hamming distance $d(x_i,x_{j})$ and
		if $d(x_i,x_{j}) \leq d_I$ flip one of the unequal bits chosen at random for $x_i$ (opinion exchange);
	\end{enumerate}
	\item
	the termination of this process as soon no more exchange is possible.
\end{enumerate}

By applying the rules repeatedly, the process converges to a stable opinion profile in which every two agents either agree in all the issues or their disagreement is larger than $d_I$.
Depending on $d_I$, different behaviour of the population is observed: low values result in a state of highly fragmented opinions and higher values yield consensus.
A precise study of the opinion distribution in the frozen state is presented in Ref.~\cite{Banisch2009}.
In the present work, we concentrate on the opinion profiles before freezing in a stable configuration.
Model parameters are chosen in order to eventually lead to a global (quasi--)consensus profile, while requiring a relatively long time to reach the absorbing state (i.e., $k = 20, d_I = 5$ and $200 \leq N \leq 4000$).

\subsection{Artificial Elections and the Transient}

\begin{figure}[htbp]
	\centering
		\includegraphics[width=1.0\linewidth]{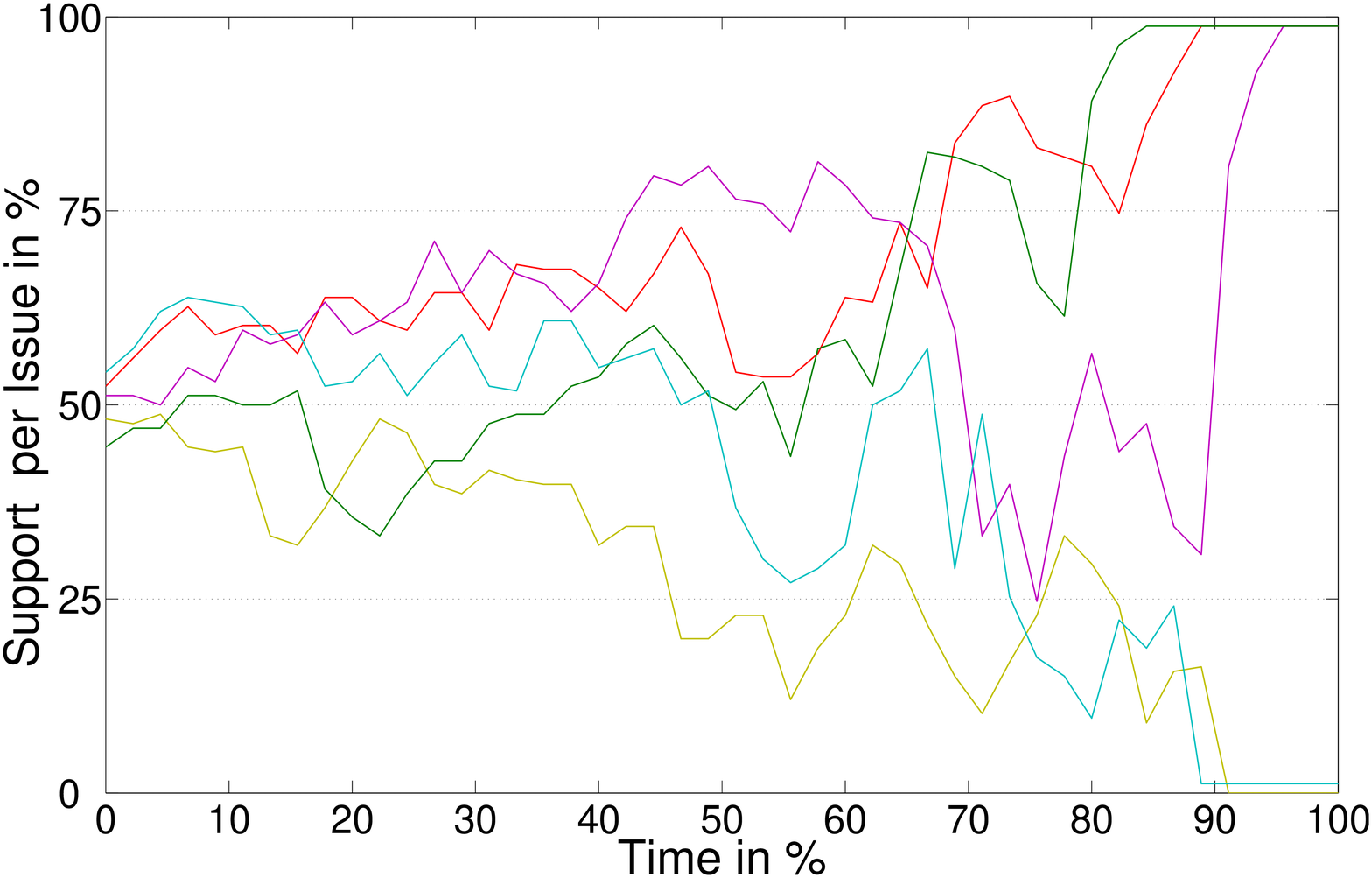}
	\caption{The dynamical phases of a model run. The curves represent the level of support regarding five issues in the binary opinion string of length $k = 20$.}
	\label{fig:Phases.PR}
\end{figure}

The dynamical evolution of the preferences is characterised by three different eras.
In the first period, called the \emph{burn--in phase} (this terminology follows the work of Laver and Sergenti, \cite{Laver2010}), preference patterns which do not deviate significantly from the random initial case are observed. 
The period after the simulation "burnt--in", we refer to as \emph{transient phase}.
The opinion structure is somewhere in between randomness and order, and the main hypothesis made in this Letter is that preference distributions comparing to real--world preference profiles have emerged.
The third and \emph{final dynamic era} is characterized by a relatively fast convergence to a stable profile with all the agents in the same state.
Fig.~\ref{fig:Phases.PR} shows the dynamical evolution of the relative support provided for five issues.

The question for the empirical confirmation experiment is whether the preference distribution observed in the transient of the model is realistic.
In order to relate opinion profiles to the electoral performance of candidates (for which the data is available \cite{FinishElections2003}) an artificial election procedure is introduced.
The election procedure is based on the well--established principle of proximity voting which assumes that a voter chooses the candidate which is closest to her/him.
Proximity voting was first proposed back in 1929 by Hotelling (\cite{Hotelling1929}) in the context of economic competition and later (in 1957) applied to the problem of candidate positioning by Downs (\cite{Downs1957}).

Initially, $Q$ random bit--strings are generated and taken to account for the policy propositions of $Q$ different candidates.
Then, a proximity voting election procedure is performed. 
The implementation of such procedure is based on the work of Ara\'{u}jo et al. in two different contexts: 
the one where consumers are driven by market--oriented innovations (\cite{Araujo2009}) and another where workers compete for jobs in a labour market (\cite{Araujo2008}).
In the procedure, the "best" candidate string is determined for each agent.
For this purpose, the Hamming distances to all the $Q$ candidate strings are computed and compared to each other.
An agent chooses that candidate string with which (s)he has the most bits in common (largest matching).
If the largest matching value of an agent is obtained with two or more candidates, we throw a fair coin (dice).
In this way the number of votes received by the $Q$ candidates is determined.

\section{Results}
\subsection{Distribution of Votes}

\begin{figure}[htbp]
	\centering
		\includegraphics[width=1.0\linewidth]{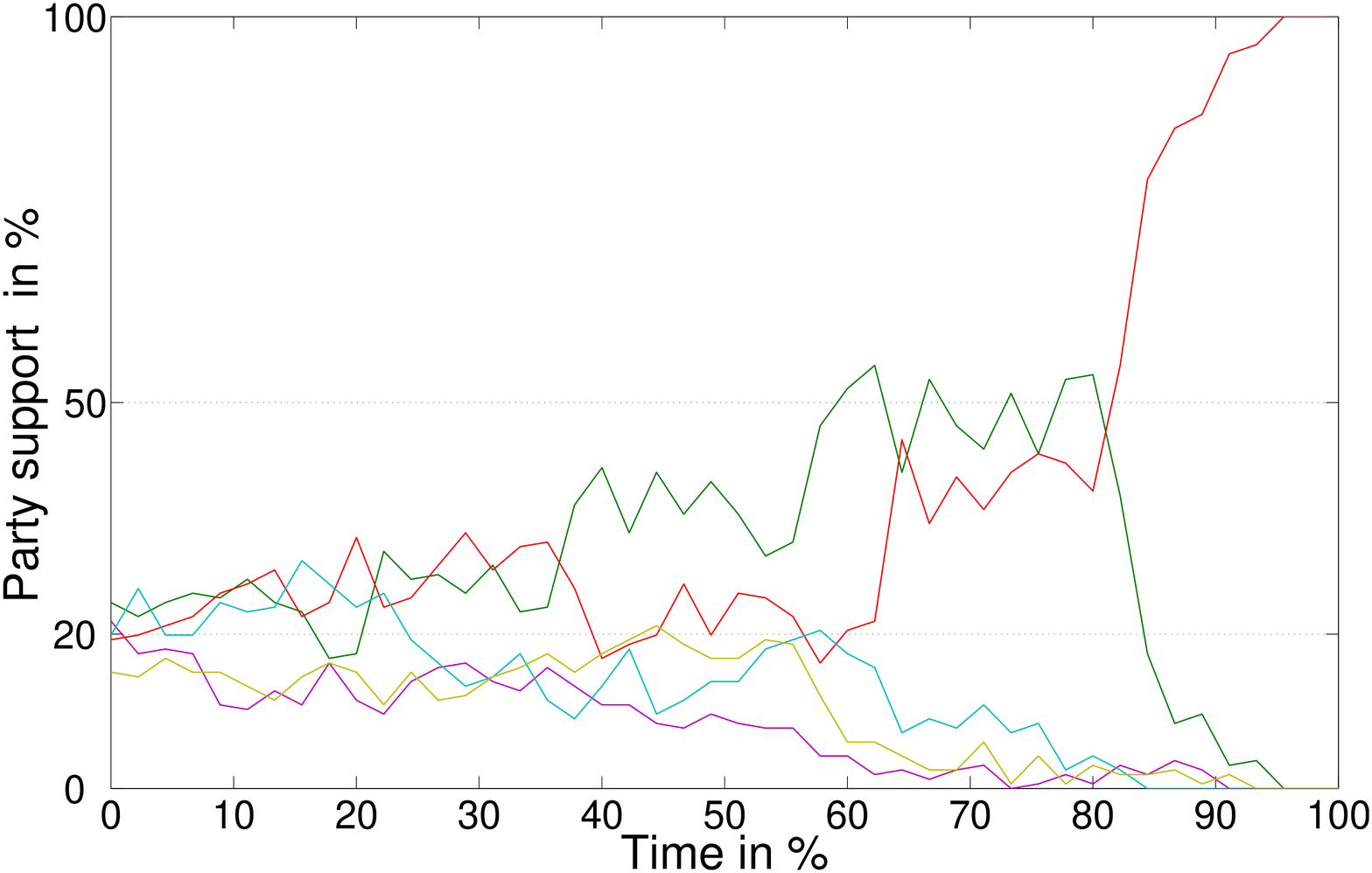}
	\caption{The dynamical phases of a model run. The curves represent a repeated voting process in which agents can choose from five different candidates with random policy positions.}
	\label{fig:Phases.ES}
\end{figure}

Fig.~\ref{fig:Phases.ES} shows a typical time--evolution of a repeated election process for five different candidates ($Q = 5$).
If five candidates apply for the votes in a population with random preferences every one of them gets approximately 20\% of the votes.
In the \emph{burn--in phase}, all the candidate support levels are near this theoretical result for random opinion states.
Only after the simulation evolved in time, do candidates perform significantly better than that (and therefore others do worse).
As the simulation continues, the population converges to a consensus configuration and all the voters eventually vote for the same candidate.
All the votes are received by the candidate which is closest to the consensus opinion string.

\begin{figure}[htbp]
	\centering
		\includegraphics[width=1.0\linewidth]{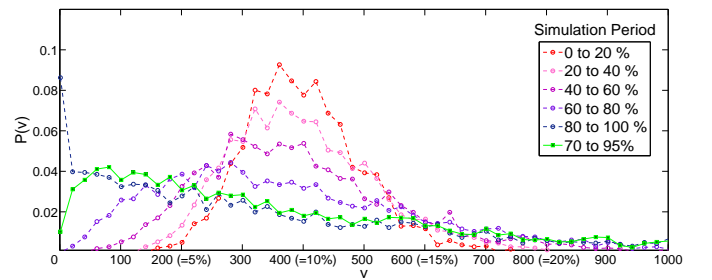}
	\caption{Voting behaviour of a population of $N=4000$ agents which choose $Q = 10$ candidates during different time periods.}
	\label{fig:Votes.Periods}
	\vspace{-5pt}
\end{figure}

We observe in Fig.~\ref{fig:Phases.ES} that for quite a long period (from 60 to 80\%) the five--candidate example yields a pattern with two major parties and three minor ones.
We currently have a similar situation in Germany. 
Though this is a first indication of realistic voting patterns, only a statistical comparison based on a larger number of elections can confirm that realistic voting behaviour is observed in the election scenario.
Therefore, a systematic computational experiment has been performed, in which a series of artificial elections is run on (a series of) evolving agent populations.
During the opinion simulation, elections are performed repeatedly after a certain number of iterations has passed.
To avoid that particular candidate positions of a certain random set--up correlate (in the sense that some strings are closer together than others) and that such positional correlations affect the statistics of the electoral performance, new random candidate strings are assigned before each election.\footnote{Note that we did not do so in Fig.~\ref{fig:Votes.Periods} and therefore a continuous electoral performance evolution is observed.}

Fig.~\ref{fig:Votes.Periods} shows how the vote distribution behaves during different periods of the model for a population of $N=4000$ agents which choose $Q = 10$ candidates.
During the \emph{burn--in phase} in the beginning of the simulation the vote distribution is similar to the normal distribution (with $\mu = \frac{Q}{N} = 400, \sigma = 100$).
In the last period many candidates have zero votes for the benefit of a single candidate which gains the support of almost all the voters.
Both cases represent unrealistic situations.
More realistic voting patterns are observed in a period of 70 to 95\% of the simulation time.
There are only very few cases of zero votes (which is more realistic as in real elections at least the candidate votes for himself).
And also candidates that arrive to be supported by 20 up to 50\% of the population are observed with reasonable frequency.
Note that only one specific electoral set--up is considered here in which 10 candidates compete for the votes of 4000 people.
This is not suited for a comparison to real proportional elections which consist of a series of heterogeneous electoral settings, but it serves as an identification of that period in the transient which should be considered in the statistical comparison.

\subsection{Statistical Comparison to Real Elections}

For this comparison a series of election experiments as described above has been performed with differing candidate numbers (from 5 to 30) and number of voters (from 200 to 4000).
To be able to compare these different voting environments, a re--scaling as proposed by Fortunato and Castellano in \cite{Fortunato2007} is applied.
In this normalisation, the number of votes ($v$) is multiplied by the number of candidates ($Q$) and divided by the number of voters ($N$) and the distribution of the function $F(\frac{vQ}{N})$ is considered.
This re--scaling is applied to the results of the repeated artificial elections on the one hand and to the results of the 2003 Finish elections on the other.
The Finland 2003 election data is available under~\cite{FinishElections2003} (170 voting sets on the whole).
\begin{figure}[htbp]
\vspace{-11pt}
	\centering
		\includegraphics[width=1.00\linewidth]{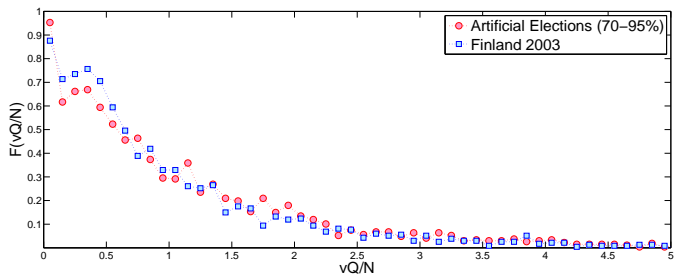}
	\caption{Comparison of the 2003 Finland elections to the results of a series of artificial elections performed on transient opinion profiles.}
	\label{fig:Votes.ComparisonFinland}
\end{figure}

The result which is shown in Fig.~\ref{fig:Votes.ComparisonFinland} is unambiguous.
The voting behaviour in the proportional elections in Finland is reproduced by elections performed on the transient opinion profiles as they form during the iteration of the opinion exchange model.
As the Finland data shares this distributional properties with other proportional elections (\cite{Fortunato2007}), this provides a strong indication that important aspects of real preference dynamics are captured by the model opinion exchange introduced in Ref.~\cite{Banisch2009}.

\section{Conclusions}

We presented an empirical confirmation experiment which allows to relate vector opinion models to election data.
This is achieved by an artificial election procedure based on the well--established principle of proximity voting which is run on transient opinion profiles.
The statistical comparison shows that preference distributions can be observed in the run of the opinion model that relate to preference distributions in real societies.
The voting behaviour in proportional elections is reproduced. 
The statistical agreement with the Finland 2003 elections is remarkable.

For these reasons, the model provides an alternative microscopic explanation for the universal voting pattern found in \cite{Fortunato2007}.
While their spreading model is very simple, the model used here has shown suitable also for the generation of realistic communication networks (\cite{Banisch2009}) so that a link to reality is provided in different domains.
We envision that future opinion studies will be more rigorously tested against empirical data and hope that the confirmation experiment introduced here will assist this development.

\section{Acknowledgements}
We acknowledge the financial support of the "Fundação para a Ciência e a Tecnologia" (FCT) Portugal, Project PDCT/EGE/60193/2004.


\end{document}